\documentclass[usenatbib]{mn2e}

\usepackage{times}
\usepackage{graphicx}
\usepackage{amssymb}

\begin{document}

\title{Recent collisional jet from a primitive asteroid}

\author[B.\ Novakovi\'c et al.]{Bojan~Novakovi\'c$^1$\thanks{E-mail:
bojan@matf.bg.ac.rs (BN)},
Aldo Dell'Oro$^2$, Alberto Cellino$^3$ and Zoran~Kne\v{z}evi\'c$^4$ \\
$^1$ Department of Astronomy, Faculty of Mathematics, University of Belgrade,
Studenski trg 16, 11000 Belgrade, Serbia \\
$^2$ INAF, Osservatorio Astrofisico di Arcetri, Largo Enrico Fermi 5, 50125 Firenze, Italy \\
$^3$ INAF, Osservatorio Astronomico di Torino, strada Osservatorio 20, 10025 Pino Torinese, Italy \\
$^4$ Astronomical Observatory, Volgina 7, 110~60 Belgrade 38, Serbia}

\maketitle \begin{abstract}
Here we show an example of a young asteroid cluster located in a dynamically stable
region, which was produced by partial disruption of a primitive body about 30 km in size. We
estimate its age to be only $1.9\pm0.3$ Myr, thus its post-impact evolution should have been very limited.
The large difference in  size between the largest object and the other cluster
members means that this was a cratering event. The parent body had a
large orbital inclination, and was subject to collisions with
typical impact speeds higher by a factor of $2$ than in the most
common situations encountered in the main belt. For the first time
we have at disposal the observable outcome of a very recent event
to study high-speed collisions involving primitive asteroids,
providing very useful constraints to numerical simulations of these
events and to laboratory experiments.
\end{abstract}

\begin{keywords}
celestial mechanics, minor planets, asteroids, methods: numerical
\end{keywords}


\section{Introduction}
\label{}

The asteroid population, being steadily subject to a process of collisional
evolution \citep{davis1989,bottke2005,morby2009,asphaug2009}, provides excellent
possibilities to study physics of collisional events. Asteroid families, which
are believed to originate from catastrophic disruption of single parent
bodies \citep{Zappala2002}, are, almost one century since the pioneering work 
by \citet{hirayama1918}, still an attractive and challenging subject.
They provide a key to our understanding of the collisional
history of the main asteroid belt \citep{bottke2005,cellino2009}, outcomes of disruption events
over a size range inaccessible to laboratory experiments \citep{michel2003,durda2007,asphaug2010},
clues on the mineralogical structure of their parent bodies \citep{cellino2002},
the role of space weathering effects \citep{Nesvorny2005,vernazza2009} and to many other subjects.

So far, ejecta from a few tens of large-scale collisions has been discovered
across the main asteroid belt \citep[e.g.][]{Zappala1995,MD2005,Nesvorny2005}.
In terms of their estimated ages, most families identified so far are fairly old
and have had enough time to evolve significantly since the epoch of their formation
as a consequence of (i) chaotic diffusion \citep{nes2002flora,Novakovic2010b},
(ii) semi-major axis drift due to Yarkovsky effect \citep{farinella1999,bottke2001},
(iii) secondary collisions \citep{Marzari1999,bottke2005},
(iv) non-destructive collisions \citep{aldo2007} and/or (v) diffusion due to close encounters
with massive asteroids \citep{carruba2003,Novakovic2010c,Delisle2012}.

In this respect, little altered recently born families may provide more direct information about the physics
of break-up events. Evidence of recent collisions in the asteroid belt have
been reported in the last decade and our knowledge about young asteroid families has been
increased significantly \citep{Nesvorny2002,datura2006,nes2006}. Most of these
groups are formed by asteroids belonging to the $S$ taxonomic class. There are, however,
several important differences among the $S$ and $C$-type asteroids. The objects belonging to former class
are thought to have experienced some thermal evolution since the time of their formation, and
it is, for example, known that space weathering processes are different for these two classes of objects \citep[e.g.][]{gaffey2010}.
Also, numerical simulations show that the outcomes of collisional events are dependent
on internal structure of the parent body \citep{jutzi2009}. Because of these reasons it is necessary
to identify also young $C$-class families in dynamically stable regions, because a few such groups
are already known, but none of these is well suited to extract reliable enough information.
Two $C$-type families, namely Veritas and Theobalda, about 8.3 and 6.7 Myr old respectively, are both
located in dynamically unstable region \citep{nes2003,Novakovic2010a}. Thus, despite their young ages,
these families evolved significantly since post-impact situation. Most of the asteroids belonging to
Beagle family \citep{Nesvorny2008}, which is probably less than 10 Myr old, are located in
dynamically relatively stable region. However, this group is embedded in the large Themis family making
distinction between the real members of the group and background objects very difficult.
Finally, the youngest known group that might be formed by $C$-type asteroids is Emilkowalski cluster,
which is only $220\pm30$ kyr old \citep{nes2006}. However, it seems to be rather an $X$- than $C$-type group
because albedos of its members are much higher than expected for $C$-type objects. For example,
geometric albedo of asteroid (14627) Emilkowalski is $0.2013\pm0.0170$ \citep{wise}.

Thus, it is of extreme importance to identify young families, that belong to the most primitive $C$
class, that do not suffer from above mentioned problems.
We have found the first example of this kind to be the Lorre cluster, recently
discovered by \citet{Novakovic2011}. According to existing color
data its largest member, (5438)~Lorre, is a primitive carbonaceous $C$-class
asteroid, which may contain organic materials.
Moreover, the members of this cluster are located in dynamical stable region and very tightly
packed in the space of proper orbital elements \citep{Kne2003}, suggesting
a likely young age. Therefore, its post-impact evolution should have been very limited.
This makes it a very promising candidate for different
possible studies. Two crucial prerequisites for these studies are an accurate
identification of its members, and a reliable estimation of its age.
These are the questions we address here.

\section{Lorre cluster}
\subsection{Membership}
\label{s:members}

A dynamical criterion for family membership is based on distances among the objects
in the space of proper orbital elements: semi-major axis ($a_{p}$), eccentricity
($e_{p}$), and inclination ($i_{p}$). Usually, for this purpose the
hierarchical clustering method (HCM) and 'standard' metric ($d$)
are used \citep{Zappala1990,Zappala1994}. This metric is defined as
\begin{equation}
d = n a_{p}\sqrt{\frac{5}{4}(\frac{\delta a_{p}}{a_{p}})^{2}+2(\delta e_{p})^{2}+2(\delta sin(i_{p}))^{2}}
\label{eq:metric}
\end{equation}
where $na_{p}$ is the heliocentric velocity of an asteroid on a
circular orbit having the semi-major axis $a_{p}$. $\delta
a_{p}$~=~$a_{p_{1}}-a_{p_{2}}$, $\delta
e_{p}$~=~$e_{p_{1}}-e_{p_{2}}$, and $\delta
\sin(i_{p})$~=~$\sin(i_{p_{1}})-\sin(i_{p_{2}})$, where the indexes (1) and (2)
denote the two bodies under consideration.
The HCM connects all objects whose mutual
distances (expressed in meters per second) are below a threshold value ($d_{c}$).

\begin{figure}
\includegraphics[angle=-90,scale=.33]{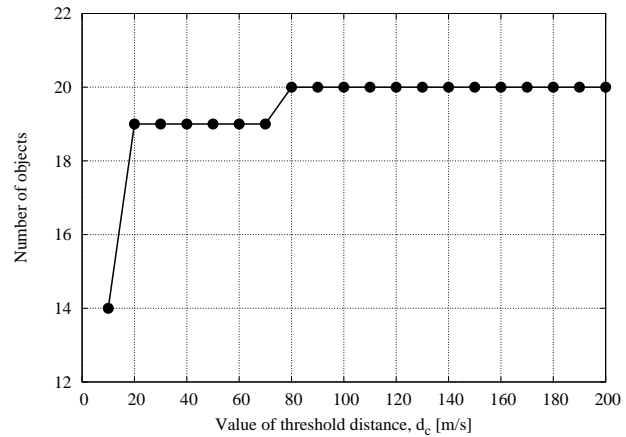}
\caption{Number of asteroids associated with Lorre cluster as a
function of cut-off distance $d_{c}$. Unusually tightly
packed members and sharp distinction of the cluster from background population are two
main characteristics.}
\label{f:nfv}
\end{figure}

Following the method described in \citet{Kne2000} we calculated synthetic proper elements for
148 asteroids located in a region somewhat wider than that occupied by the cluster.
This region covers the following ranges in the osculating orbital elements: 2.738 $< a <$ 2.758 au, 
0.13 $< e <$ 0.39 and 23 $< i <$ 31$^\circ$. The number of asteroids includes numbered, multi- and 
single-opposition objects\footnote{Although the orbits of 
single-opposition objects are less reliably known, we used them as well in order to
find as many cluster members as possible.}, found in the recent version of
catalogs of osculating elements retrieved from the AstDys web
page.\footnote{Asteroids Dynamic Site: http://hamilton.dm.unipi.it/astdys2/}
Then, we applied the HCM to this set of proper elements, and we
analyzed the number of dynamically linked objects identified at
different mutual distances (Fig.~\ref{f:nfv}). In particular, this was
done by changing $d_{c}$ from 10 to 200 m~s$^{-1}$ at discrete steps
of 10 m~s$^{-1}$. At the lowest tested value of $d_{c} = 10$ m~s$^{-1}$
the HCM links 14 asteroids with Lorre, while the number of members
raises to 19 for 20~m s$^{-1}$. The number of dynamically
associated members remains a constant until $80$ m~s$^{-1}$, when one
body, asteroid 2006 AX$_{67}$ is added. Later on, no
additional body is linked to the cluster, even for the largest used
value of $d_{c} = 200$ m~s$^{-1}$.

From these results we can draw three basic conclusions: 
(i) the cluster is extremely compact and very well separated from the
background population; (ii) the nominal membership of the cluster is
best characterized at $d_{c} = 20$ m~s$^{-1}$; (iii)
the asteroid connected with the group at 80 m~s$^{-1}$ is likely a
close background object. Thus, the Lorre cluster has 19 currently known members (Table~\ref{t:lorre_prop}).
These asteroids are very tightly packed with mutual distances
significantly smaller than in the cases of typical families in the main asteroid belt.

\begin{table}
\scriptsize
 \centering
  \caption{Proper orbital elements of the asteroids belonging to the Lorre cluster.
In columns are given: semi-major axis ($a_{p}$), eccentricity ($e_{p}$), sine of inclination ($sin(i_{p})$),
mean motion ($n$), frequency of the longitude of perihelion ($g$) and frequency of the longitude of node ($s$).} \label{t:lorre_prop}
  \begin{tabular}{lcccccr}
\hline
Asteroid & $a_{p}$ [au] & $e_{p}$ & $sin(i_{p})$ & $n$ [\degr/ yr] & $g$ [\arcsec/ yr] & $s$ [\arcsec/ yr]  \\
\hline
5438             & 2.74732 & 0.26290 & 0.47230 & 79.0466 & 9.4486  & -49.7809  \\
208099           & 2.74694 & 0.26314 & 0.47241 & 79.0630 & 9.4207  & -49.7557  \\
2001 RF$_{42}$   & 2.74427 & 0.26321 & 0.47176 & 79.1781 & 9.5380  & -49.7308  \\
2001 XF$_{167}$  & 2.74718 & 0.26314 & 0.47253 & 79.0532 & 9.3965  & -49.7490  \\
2003 BW$_{5}$    & 2.74796 & 0.26294 & 0.47198 & 79.0210 & 9.4916  & -49.8129  \\
2003 YY$_{120}$  & 2.74671 & 0.26342 & 0.47212 & 79.0732 & 9.4846  & -49.8060  \\
2005 YD$_{18}$   & 2.74788 & 0.26313 & 0.47246 & 79.0204 & 9.4144  & -49.7840  \\
2006 AL$_{16}$   & 2.74636 & 0.26342 & 0.47211 & 79.0883 & 9.4880  & -49.7946  \\
2006 RM$_{98}$   & 2.74263 & 0.26276 & 0.47201 & 79.2495 & 9.4847  & -49.6157  \\
2007 BJ$_{62}$   & 2.74626 & 0.26338 & 0.47204 & 79.0927 & 9.4968  & -49.7919  \\
2008 AD$_{104}$  & 2.74722 & 0.26308 & 0.47240 & 79.0511 & 9.4276  & -49.7716  \\
2010 CG$_{176}$  & 2.74536 & 0.26292 & 0.47195 & 79.1313 & 9.4935  & -49.7350  \\
2011 FQ$_{151}$  & 2.74521 & 0.26299 & 0.47196 & 79.1377 & 9.4915  & -49.7297  \\
\hline
2010 AX$_{32}$   & 2.74668 & 0.26382 & 0.47227 & 79.0744 & 9.4695  & -49.8253  \\
2006 VZ$_{122}$  & 2.74783 & 0.26342 & 0.47239 & 79.0216 & 9.4505  & -49.8372  \\
2008 BB$_{10}$   & 2.74631 & 0.26348 & 0.47220 & 79.0899 & 9.4758  & -49.7979  \\
2008 DE$_{8}$    & 2.74490 & 0.26318 & 0.47193 & 79.1514 & 9.5111  & -49.7482  \\
2010 EW$_{42}$   & 2.74544 & 0.26333 & 0.47198 & 79.1277 & 9.4987  & -49.7634  \\
2010 EJ$_{81}$   & 2.74233 & 0.26344 & 0.47210 & 79.2627 & 9.4890  & -49.6575  \\
\hline
\end{tabular}
\end{table}

\begin{figure}
\includegraphics[angle=-90,scale=.33]{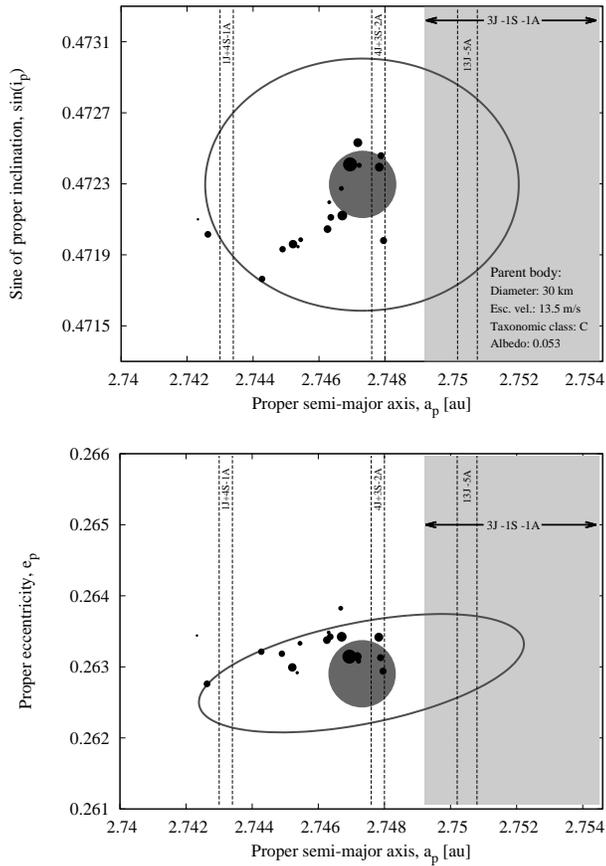}
\caption{The Lorre cluster in the space of proper elements. The size
of each symbol is proportional to the diameter of the body. The
superimposed ellipses represent equivelocity curves, computed
according to the Gaussian equations \citep{nes2006}. These ellipses
are obtained assuming a velocity change
$\Delta v = 15$ m~s$^{-1}$, argument of perihelion $\omega =
90 \degr$, and true anomaly $f = 90 \degr$. The ellipses are shown to illustrate
distribution of the fragments in the case of isotropic ejection field;
however, it is easy to see that the ejection velocity field of the Lorre cluster was highly asymmetric,
a nice example of what one should expect to be the outcome of a
cratering event \citep{voknes2011}. The locations of the relevant
mean motion resonances are denoted with the vertical dashed lines, except
in the case of 3J$-$1S$-$1A resonance that is marked with the gray-shaded region.
In the top panel basic information about parent body are also given.}
\label{f:aei}
\end{figure}

\subsection{Dynamics}
\label{s:dynamics}

The orbits of the asteroids belonging to the Lorre cluster are characterized by the moderate
eccentricities ($e_{p} \approx 0.26$) and high inclinations ($i_{p} \approx 28 \degr$), but
the region occupied by these asteroids is not under influence of any of the strong mean motion
or secular resonances. Thus, despite their orbital characteristics, these asteroids are mostly
stable. Still, there are a few mean motion resonances (MMRs), present in the region, whose
influence should not be neglected. The most powerful is a three-body\footnote{Three-body mean 
motion resonances are comensurabilities between
the mean motions of Jupiter, Saturn and asteroid \citep{Nesvorny1998}. They are characterized
by the relation $m_{J} \dot{\lambda}_{J} + m_{S} \dot{\lambda}_{S} + m \dot{\lambda} \sim 0$,
where $\dot{\lambda}_{J}$, $\dot{\lambda}_{S}$ and $\dot{\lambda}$ denote mean motions of
Jupiter, Saturn, and asteroid respectively, while $m_{J}$, $m_{S}$ and $m$ are integers.} MMR 3J$-$1S$-$1A 
located at 2.752~au. A somewhat less significant but still relevant are another two 3-body
MMRs, namely 1J+4S$-$1A and 4J+3S$-$2A (see Fig.~\ref{f:aei}). Finally, 13J/5A 2-body MMR,
among Jupiter and asteroid, is present in the region as well.

To better understand the strength of these resonances and their possible influence on the
dynamical stability we have determined Lyapunov times ($T_{lyap}$) for all members of the
Lorre cluster. This was done according to the method proposed by \citet{MilNob92} and
within the framework of several different dynamical models.

As for most of the purposes, in this part of the main asteroid belt, dynamical model with
four major planets (from Jupiter to Neptune) is accurate enough, we first used this model
to estimate Lyapunov times.
The obtained values of $T_{lyap}$ are in most cases longer than 100~kyr. A few
exceptions include objects located around $a_{p} = 2.7478$~au, that are probably trapped
inside the 4J+3S$-$2A resonance. However, even Lyapunov times of these objects are
not shorter than about $30$~kyr (Table~\ref{t:lorre_tl}).

When dynamical model with seven planets, from Venus to Neptune, is used, the estimated
Lyapunov times are noticeably shorter (Table~\ref{t:lorre_tl}), meaning that this
model should be used for asteroids located in the region of Lorre cluster. The reasons
for the important difference among the results obtained with 4- and 7-planets are relatively
large orbital eccentricities and inclinations of these objects. Still, according to
this result most of the Lorre cluster members are reasonably stable, with the only one
possible exception, asteroid 2003 BW$_{5}$.

\begin{table}
\scriptsize
 \centering
  \caption{Lyapunov times of Lorre cluster members derived using different dynamical models.} \label{t:lorre_tl}
  \begin{tabular}{lrrrrrrr}
\hline

Asteroid & 4 pla & 7 pla &  7 pla + Ceres & 7 pla + CV & 7 pla + CPV \\
\hline
5438            & 107.6 & 41.7 & 35.0 & 27.2 & 26.0  \\
208099          & 128.1 & 75.4 & 50.6 & 32.5 & 30.7  \\
2001$RF_{42}$   & 290.7 & 57.7 & 36.0 & 29.0 & 16.1  \\
2001$XF_{167}$  &  82.1 & 48.2 & 31.4 & 35.1 & 30.6  \\
2003$BW_{5}$    &  41.2 &  6.7 & 21.8 &  7.9 & 20.5  \\
2003$YY_{120}$  & 534.8 &203.3 & 39.9 & 30.0 & 32.7  \\
2005$YD_{18}$   &  32.5 & 38.9 & 29.3 & 25.1 & 20.9  \\
2006$AL_{16}$   & 304.0 &162.1 & 41.5 & 35.8 & 31.4  \\
2006$RM_{98}$   & 226.2 & 22.1 & 14.0 & 19.2 & 19.7  \\
2007$BJ_{62}$   & 289.1 &219.3 & 39.7 & 34.9 & 35.1  \\
2008$AD_{104}$  & 125.6 & 41.6 & 37.8 & 28.6 & 26.0  \\
2010$CG_{176}$  &7142.9 & 75.1 & 37.4 & 36.9 & 23.0  \\
2011$FQ_{151}$  & 505.5 &115.1 & 34.4 & 27.1 & 24.7  \\
2010$AX_{32}$   &1960.0 &198.4 & 40.9 & 35.2 & 31.0  \\
2006$VZ_{122}$  &  35.4 & 31.8 & 27.5 & 25.1 & 11.5  \\
2008$BB_{10}$   & 552.9 &202.0 & 43.6 & 35.6 & 29.7  \\
2008$DE_{8}$    &4000.5 & 83.8 & 37.6 & 35.3 & 31.3  \\
2010$EW_{42}$   &2381.0 & 76.0 & 44.6 & 33.7 & 28.8  \\
2010$EJ_{81}$   & 746.6 & 97.9 & 42.3 & 34.2 & 34.1  \\
\hline
\end{tabular}
\end{table}

Recently, \cite{laskar2011} showed that close encounters with massive asteroids may induce
chaos in their and in the motion of other asteroids. To check whether or not this is the case for
Lorre cluster members, we have also calculated Lyapunov times using dynamical models that
include some of the most massive asteroids, Ceres, Pallas and Vesta.\footnote{For this purpose,
the masses of Ceres, Vesta and Pallas are set to $4.757$, $1.300$ and $1.010\times 10^{-10}M_{\sun}$
respectively \citep{mike2010,baer2011}. These masses are results of the latest calculations performed by
means of the improved methodology. A preliminary estimation of Vesta's mass provided by Dawn mission
(http://dawn.jpl.nasa.gov/mission/) perfectly match
the results from these two papers, for this object. Due to these reasons we chose to use
these values, despite being slightly smaller than those used by \cite{laskar2011}.}

Our result generally confirms that obtained by \citet{laskar2011}. Lyapunov times
become, on average, shorter when the massive asteroids are included in the dynamical model.
There are, however, a few asteroids whose motion seem to be more stable in this case, and
their values of $T_{lyap}$ are longer, than those obtained in the model with 7-planets only.
An illustrative example is the only possibly unstable object among the currently known members of the cluster,
the asteroid 2003 BW$_{5}$. Its estimated $T_{lyap}$ is only $7$ kyr in the dynamical model
with 7 planets, but rises to $22$ kyr when Ceres is added to the dynamical model.\footnote{This 
is not a surprise because an estimation of Lyapunov times, even for moderately chaotic orbits, 
is probabilistic, thus not highly reliable and should be interpreted with a care\citep{Kne2005}.}
Thus, although influence of the massive asteroids on the motion of asteroids belonging to 
the Lorre cluster is undoubtedly confirmed, its resulting effect may vary
from case to case. 

The conclusion that we can draw from derived values of Lyapunov times is that orbits of the
Lorre cluster members are neither perfectly stable nor strongly chaotic.

In terms of a possible post-impact dynamical evolution of the
cluster even week chaos may be important. Hence, to explore this possibility and
to assess a jet like shape of the cluster, we checked stability of the proper
eccentricity and inclination of asteroids belonging to the Lorre cluster.
Using the numerical integrations of cluster members performed
in the dynamical model that includes seven planets (from Venus to Neptune) and three
most massive asteroids (Ceres, Pallas and Vesta), we estimated average evolution
rates of eccentricity and sine of inclinations to be $1 \times 10^{-4}$ and $5 \times 10^{-5}$
per one million years respectively. These are slow changes that do not seem to be able
to significantly change overall structure of the cluster. Actually, as we found the Lorre
cluster to be only about 1.9~Myr old (see Section~\ref{s:age}),
over its lifetime expected changes of eccentricity and sine of inclination are only about
$2 \times 10^{-4}$ and $1 \times 10^{-4}$ respectively. By comparing these values with
the scales of y-axes in Fig.~\ref{f:aei} we concluded that dynamical evolution
is negligible.

Looking at Fig.~\ref{f:aei} it can be easily realized that distribution of the Lorre
cluster members is highly asymmetric with respect to the largest member, asteroid (5438)~Lorre.
To understand the reasons for this, we extend our dynamical analysis to
the region surrounding the cluster. The dynamical instability starts
to increase for values of semi-major axis larger than 2.748~au. The inner border
of the powerful 3J$-$1S$-$1A MMR is found at about
2.749~au. However, using numerical integrations of $100$ massless test
particles we have verified that this instability cannot explain the
absence of cluster members in the 2.748 - 2.754~au range (see
Fig.~\ref{f:aei}). Although, over a time scale of 2 Myr, many particles
interact with the 3J$-$1S$-$1A resonance, they still remain close
enough to be recognized by the HCM.

Available evidence suggests therefore that the observed asymmetry of
the family is mostly a consequence of the original ejection velocity
field of the fragments, rather than dynamical post-impact evolution.
Thus, this cluster still keeps memory of the original ejection
velocity field, a useful input to study impact physics.

\subsection{Age}
\label{s:age}

The most accurate method known so far to estimate the age of a young asteroid
family is to integrate the orbits of its members backwards in time and to identify
the epoch of their convergence \citep{Nesvorny2002,nes2003}.
However, this method can be applied only to the objects on stable orbits.
As we showed in Section~\ref{s:dynamics} that the orbits of the Lorre cluster members
are not perfectly stable an application of the backward integration method (BIM)
is not so straightforward. To overcome this problem we turn to a statistical approach
based on the BIM \citep{nes2006,voknes2011}. Instead of orbits of nominal members we
used a number of cloned, statistically equivalent, orbits. In this way we were able
to characterize the age of the Lorre cluster in a statistical sense.

More in particular, we took into account the current orbital uncertainties of the
nominal orbits and different possible evolutions of the orbital semi-major axes due to the Yarkovsky
effect. For each nominal member of the cluster, except for asteroid (5438) Lorre,
we produced a set of $10$ orbital clones. These clones are drawn from $3\sigma$ interval of their
formal uncertainties\footnote{For single-opposition objects we used the following values for all objects:
$\sigma_{a}=2.0 \times 10^{-5}$~au, $\sigma_{e}=3.0 \times 10^{-5}$, $\sigma_{i}=1\degr0 \times 10^{-4}$,
$\sigma_{\Omega}=2\fdg0 \times 10^{-4}$, $\sigma_{\omega}=3\fdg5 \times 10^{-4}$ and $\sigma_{M}=5\degr0 \times 10^{-3}$.}
listed in Table~\ref{t:lorre_osc}, assuming Gaussian distribution. Then, for each of the orbit clones we generated
$10$ different 'yarko' clones uniformly distributed over the interval stretching from zero to the maximum
expected drift due to the Yarkovsky force \citep{bottke2001}.
The maximum drift in the proper semi-major due to the Yarkovsky force $(da/dt)_{max}$ for each object
is obtained assuming thermal parameters appropriate for $C$-type asteroids \citep{broz2008}.
In this way a total of $100$ statistically
equivalent clones were assigned to each member. Clones are not used for asteroid (5438) Lorre itself
because on one hand its orbit is very well determined, while on the other
hand it is large enough (see Table~\ref{t:lorre_phy}) that Yarkovsky effect on its orbit can be safely
neglected.

The orbits of all clones were numerically integrated backward in time for 10~Myr using
the Orbit9 software. These integrations were performed
within the framework of a dynamical model that includes seven
planets, from Venus to Neptune, as perturbing bodies, and
accounts also for the Yarkovsky effect.\footnote{For simplicity, the Yarkovsky effect is included in the model as
a constant secular drift (inwards or outwards) of the semi-major axis. This approximation
seems appropriate for our purpose to characterize the age of Lorre cluster in a statistical sense.}
To account for the indirect effect of Mercury, its mass is added to the mass of the Sun and the 
barycentric correction is applied to the initial conditions.

The age of the cluster was estimated by randomly selecting one clone
for each member and determining the age for that particular
combination of clones as the minimum of the function \citep{voknes2011}:
\begin{equation}
\Delta V = n a \sqrt{(sin(i) \Delta \Omega)^{2} + 0.5(e \Delta \varpi)^{2}}
\label{eq:fun}
\end{equation}
where $na \approx$ 18 km~s$^{-1}$ is the mean orbital speed of the
asteroids in the cluster, and $\Delta \Omega$ and $\Delta \varpi$
are the dispersions of the longitude of node and the longitude of
perihelion, respectively.

\begin{figure}
\includegraphics[angle=-90,scale=.39]{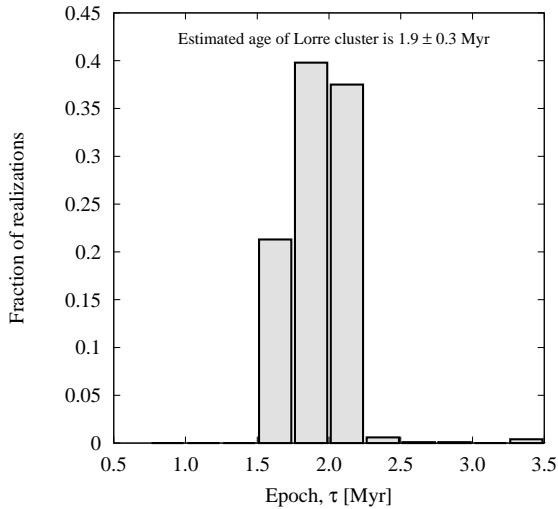}
\caption{The histogram of possible ages of Lorre cluster. It is constructed using $10^{6}$ different
combinations of clones (see text).
The age of cluster derived from these values is $1.9\pm0.3$~Myr.}
\label{f:his}
\end{figure}

The obtained results are shown in Fig.~\ref{f:his}. The age of the Lorre
cluster turns out to be $1.9\pm0.3$~Myr. The estimated error comes mainly
from the assumed orbital uncertainties of single-opposition asteroids.
Nevertheless, the result is robust and undoubtedly confirms that the
Lorre cluster is very recent.

\begin{table*}
\begin{minipage}{165mm}
\scriptsize
 \centering
\caption{The osculating orbital elements along with their formal uncertainties of Lorre cluster members at epoch 56000.0 MJD as found at AstDys.
The horizontal line separates single-opposition from multi-opposition and numbered asteroids.}
\label{t:lorre_osc}
  \begin{tabular}{lccrrrr}
\hline
Asteroid & $a$ [au] & $e$ & $i$ [\degr] &$\Omega$  [\degr] & $\omega$  [\degr] & $M$  [\degr] \\
         & $\sigma_{a}$ & $\sigma_{e}$ & $\sigma_{i}$ &$\sigma_{\Omega}$ & $\sigma_{\omega}$ & $\sigma_{M}$   \\
\hline
(5438)~Lorre             &    2.7457268384&   0.2763423275&  26.57394988& 298.51646410& 238.57467176&  53.56378415\\
                         &    0.0000000219&   0.0000001256&   0.00001231&   0.00001893&   0.00002667&   0.00002319\\
(208099)~2000 AO$_{201}$  &    2.7473300263&   0.3276837375&  24.27112923& 276.31365913& 264.96425896& 214.38044104\\
                         &    0.0000000680&   0.0000001692&   0.00002116&   0.00003503&   0.00005259&   0.00004912\\
2001 RF$_{42}$            &    2.7450585299&   0.3013219481&  26.03094469& 335.00946624& 290.68454399& 156.88949896\\
                         &    0.0000001903&   0.0000060800&   0.00017210&   0.00007168&   0.00124400&   0.00041920\\
2001 XF$_{167}$           &    2.7455657272&   0.3319144657&  24.31187210& 265.98666039& 263.95850331&  43.55344479\\
                         &    0.0000001722&   0.0000010250&   0.00005721&   0.00005738&   0.00043300&   0.00026990\\
2003 BW$_{5}$             &    2.7492826344&   0.1642237556&  29.71864289& 327.57253227& 185.45868314& 348.04765530\\
                         &    0.0000006285&   0.0000054500&   0.00018790&   0.00006704&   0.00717400&   0.00513700\\
2003 YY$_{120}$           &    2.7492437940&   0.2127232504&  28.47791027& 313.19691527& 219.29358380& 255.46889835\\
                         &    0.0000199500&   0.0000167200&   0.00028020&   0.00005237&   0.00443200&   0.00896100\\
2005 YD$_{18}$            &    2.7469026458&   0.3315685526&  24.26623885& 260.56326451& 270.39827790&  73.87404282\\
                         &    0.0000017310&   0.0000033110&   0.00008139&   0.00006975&   0.00086010&   0.00032190\\
2006 AL$_{16}$            &    2.7458145602&   0.1890859883&  29.17656654& 320.04652207& 207.87949279& 111.90710109\\
                         &    0.0000037100&   0.0000044950&   0.00011930&   0.00005157&   0.00459600&   0.00394200\\
2006 RM$_{98}$            &    2.7447515579&   0.3354039030&  25.70806090&  17.59630958& 272.87032056& 107.97789940\\
                         &    0.0000016040&   0.0000014330&   0.00009535&   0.00006296&   0.00023750&   0.00034440\\
2007 BJ$_{62}$            &    2.7447593567&   0.2247894021&  28.36497567& 331.14470720& 224.85376531&   9.52332479\\
                         &    0.0000151200&   0.0000026950&   0.00027740&   0.00004350&   0.00175900&   0.00232100\\
2008 AD$_{104}$           &    2.7496997978&   0.2884073048&  25.97423465& 292.70130574& 243.43414072& 289.50849464\\
                         &    0.0000004817&   0.0000132900&   0.00047830&   0.00007045&   0.00215700&   0.00096190\\
2010 CG$_{176}$           &    2.7442227579&   0.3251930245&  24.67267683& 325.68407481& 268.76680008& 109.02511741\\
                         &    0.0000616600&   0.0000183400&   0.00024660&   0.00006977&   0.00062410&   0.00311900\\
2011 FQ$_{151}$           &    2.7448445904&   0.3017178312&  25.91127578& 342.12860292& 250.33266440&  51.86563823\\
                         &    0.0000005331&   0.0000004595&   0.00005290&   0.00006177&   0.00018470&   0.00008633\\
\hline
2010 AX$_{32}$            &    2.7455415640&   0.1900949595&  29.19226479& 304.42259405& 204.52702785& 151.89267673\\
2006 VZ$_{122}$           &    2.7464620372&   0.3261117058&  24.64077904& 247.41989238& 280.28681378& 352.79862640\\
2008 BB$_{10}$            &    2.7471789321&   0.2118524243&  28.54630827& 321.49428079& 220.23573222& 296.43085357\\
2008 DE$_{8}$             &    2.7439824242&   0.2862965047&  26.49942006& 328.29934893& 295.69242679& 238.32412135\\
2010 EW$_{42}$            &    2.7450187057&   0.2760147066&  27.13103288& 354.38366196& 241.26515384& 116.01575514\\
2010 EJ$_{81}$            &    2.7455510601&   0.3260448227&  25.33179996& 356.59785342& 263.30030408& 108.50902398\\
\hline
\end{tabular}
\end{minipage}
\end{table*}

\subsection{Physical and spectral characteristics}

As for physical properties, the geometric albedos ($p_{v}$) have
been determined for $6$ members of the Lorre cluster
\citep{akari,wise}, with an average value of $0.053$, compatible
with $C$-class objects.

Unfortunately, little is known about the spectral reflectance
properties. To date, a spectral class has been determined only for
the largest asteroid (5438)~Lorre, which is classified as a $C$-type
\citep{taxonomy2002}.

It is interesting to note that for this same asteroid an estimate of
the rotational period $P$ is also available. According to
\citet{Behrend} $P$ is about 25 hours. This unusually long period
might be, at least partly, the result of angular momentum transfer
during the impact \citep{DB1984,cellino90,Takeda2009}, that may
produce in some cases despinning of mid-sized objects.

\begin{table}
\scriptsize
 \centering
  \caption{Different characteristics of the Lorre cluster members.} \label{t:lorre_phy}
  \begin{tabular}{lccrc}
\hline
Asteroid & $H$  &  $p_{v}\pm \sigma_{A}$ & $D\pm \sigma_{D}$ & $(da/dt)_{\rm max}$  \\
         &[mag] &   &  [km] &  [au~Myr$^{-1}$] \\

\hline
5438             &11.4  &$0.069\pm0.002$ & $30.1\pm0.4$ & $1.5\times10^{-5}$ \\
208099           &14.8  &$0.052\pm0.008$ & $6.1\pm0.1$  & $7.5\times10^{-5}$ \\
2001 RF$_{42}$   &16.5  &$0.060\pm0.024$ & $ 2.3\pm0.2$ & $2.0\times10^{-4}$ \\
2001 XF$_{167}$  &15.8  &       -        &      -       & $1.2\times10^{-4}$ \\
2003 BW$_{5}$    &16.3  &       -        &      -       & $1.4\times10^{-4}$ \\
2003 YY$_{120}$  &15.6  &$0.045\pm0.021$ & $ 4.3\pm1.2$ & $1.0\times10^{-4}$ \\
2005 YD$_{18}$   &16.3  &       -        &      -       & $1.4\times10^{-4}$ \\
2006 AL$_{16}$   &16.4  &$0.058\pm0.021$ & $ 3.0\pm0.3$ & $1.5\times10^{-4}$ \\
2006 RM$_{98}$   &16.4  &$0.036\pm0.003$ & $ 3.7\pm0.1$ & $1.2\times10^{-4}$ \\
2007 BJ$_{62}$   &16.1  &       -        &      -       & $1.3\times10^{-4}$ \\
2008 AD$_{104}$  &17.0  &       -        &      -       & $2.0\times10^{-4}$  \\
2010 CG$_{176}$  &17.9  &       -        &      -       & $3.1\times10^{-4}$  \\
2011 FQ$_{151}$  &15.9  &       -        &      -       & $1.2\times10^{-4}$  \\
2010 AX$_{32}$   &17.1  &       -        &      -       & $2.0\times10^{-4}$  \\
2006 VZ$_{122}$  &15.8  &       -        &      -       & $1.2\times10^{-4}$  \\
2008 BB$_{10}$   &17.6  &       -        &      -       & $2.7\times10^{-4}$  \\
2008 DE$_{8}$    &16.5  &       -        &      -       & $1.6\times10^{-4}$  \\
2010 EW$_{42}$   &17.1  &       -        &      -       & $2.0\times10^{-4}$  \\
2010 EJ$_{81}$   &18.6  &       -        &      -       & $4.2\times10^{-4}$  \\
\hline
\end{tabular}
\end{table}

\subsubsection{Size of the parent body}

To further characterize the event which produced the Lorre cluster
we estimated the size of the parent body. The simplest way to
achieve this goal is to estimate the volume of the parent body by
summing up the volumes of all known members, assuming a spherical
shape for all of them. For this purpose we used the available diameters
of the objects obtained by thermal radiometry observations, using WISE
data in all cases given in Table~\ref{t:lorre_prop}, but for Lorre 
itself, whose diameter is known from AKARI observations. For the objects 
lacking a size estimate (the majority of the objects in our sample), we 
derived it using the well known relation between diameter, absolute 
magnitude and albedo (see below).

We adopted for each object the nominal value of its absolute magnitude 
$H$ taken from the AstDys catalog, (these data are also listed in 
Table~\ref{t:lorre_prop}. One should be aware that the catalog values
of $H$ are known to be affected by large uncertainties for objects in
this magnitude range \citep{Muinonenetal2010}. This also affects
negatively the errors in the albedo determined by means of the thermal 
radiometry technique, and for this reason we tend to believe that the
nominal values listed in Table~\ref{t:lorre_prop} for WISE and
AKARI-derived albedos may well be quite optimistic in some cases.
For each object lacking an albedo measurement, we adopted the average 
value of $0.053$ for this family, which is based on the nominal values 
shown in Table~\ref{t:lorre_prop}). From $H$ and the albedo we can derive 
the size from the relation $\log(D) = 3.1236 -0.2\,H -0.5\log(p_V)$
where $D$ is the diameter. The obtained $D$ values range between
$1.1$ and $4.1$ km.

By summing up all the resulting volumes of the family members, we
find that the parent body was just a little larger than the largest
fragment\footnote{The difference among diameters of the parent body
and largest fragment is smaller than the uncertainties of these two
values.}, (5438)~Lorre, which has an estimated diameter
about 30~km. This conclusion does not change if we simply assume
that the parent body could not be smaller than the sum of the
sizes of the two largest family members. This is the criterion
applied by \citet{Tangaetal99}, and it is based on simple geometric
considerations, which is more suitable to treat the cases of full
parent body disruption.

The escape velocity from a surface of a $30$-km body is about 13.5
m~s$^{-1}$ (assuming a density of 1.5 g~cm$^{-3}$, typical of
$C$-class asteroids). The second largest member of the cluster,
asteroid (208099) 2000 AO$_{201}$, is about 6~km in diameter. The
cluster turns out to be therefore the outcome of a cratering event,
which was not sufficiently energetic to completely disrupt the
parent body. This result supports our conclusion that the observed
asymmetry of the cluster is likely a consequence of the original
ejection velocity field.

\subsubsection{Ejection velocity field}

The structure of the families in the space of proper elements can be
used to infer some information on the ejection velocities of the
fragments in family-forming events \citep{Zappala2002}. As we
already noted, the most important feature of the ejection velocity
field (EVF) of Lorre seems to be a high asymmetry with respect to
the location of the largest member. This, however, is not the only
peculiar characteristics of the EVF. A jet like structure is visible
in both, ($a_{p}$,$e_{p}$) and ($a_{p}$,$sin(i_{p})$), planes
(Fig.~\ref{f:aei}). This is not unexpected in the case of a
cratering event. Jetting is expected to have a chance to occur when
two objects collide at high speeds and at high incidence angles
\citep[see e.g.][]{housen2011}. Such structure is observed in both
numerical simulations and laboratory experiments \citep{yang1995},
but it has not been observed yet among real asteroid families,
mainly due to the post-impact evolution of the known groups.

\begin{table}
\scriptsize
 \centering
  \caption{Differences in velocities with respect to asteroid (5438)~Lorre.} \label{t:lorre_evf}
  \begin{tabular}{lrrrr}
\hline
Asteroid & $\Delta v_{a_{p}}$ [m/s] &  $\Delta v_{e_{p}}$ [m/s] & $\Delta v_{sin(i_{p})}$ [m/s] & $\Delta v$ [m/s] \\
\hline
5438             &  0.0 &  0.0 &   0.0   &  0.0 \\
208099           &  1.5 &  3.2 &   1.5   &  3.8 \\
2001 RF$_{42}$   & 11.7 &  4.1 &   7.1   & 14.2 \\
2001 XF$_{167}$  &  0.6 &  3.2 &   3.1   &  4.5 \\
2003 BW$_{5}$    &  2.4 &  0.5 &   4.2   &  4.9 \\
2003 YY$_{120}$  &  2.4 &  6.9 &   2.3   &  7.7 \\
2005 YD$_{18}$   &  2.1 &  3.0 &   2.1   &  4.3 \\
2006 AL$_{16}$   &  3.7 &  6.9 &   2.5   &  8.2 \\
2006 RM$_{98}$   & 17.9 &  1.9 &   3.7   & 18.4 \\
2007 BJ$_{62}$   &  4.1 &  6.3 &   3.3   &  8.2 \\
2008 AD$_{104}$  &  0.4 &  2.4 &   1.4   &  2.8 \\
2010 CG$_{176}$  &  7.5 &  0.2 &   4.7   &  8.8 \\
2011 FQ$_{151}$  &  8.1 &  1.2 &   4.5   &  9.3 \\
2010 AX$_{32}$   &  2.5 & 12.3 &   0.3   & 12.5 \\
2006 VZ$_{122}$  &  1.9 &  6.8 &   1.3   &  7.2 \\
2008 BB$_{10}$   &  3.9 &  7.7 &   1.3   &  8.7 \\
2008 DE$_{8}$    &  9.3 &  3.7 &   4.8   & 11.1 \\
2010 EW$_{42}$   &  7.2 &  5.7 &   4.1   & 10.1 \\
2010 EJ$_{81}$   & 19.1 &  7.1 &   2.6   & 20.6 \\
\hline
\end{tabular}
\end{table}

Although a detail study of the EVF is beyond the scope of this
paper, we want to emphasize here that there is a clear trend in the
velocity-size relationship. This trend is in agreement with previous
studies \citep{cellino1999} suggesting that smaller fragments are
ejected on the average with slightly higher velocities. However, the
number of known cluster members is still too small at the moment to
analyze this trend in more detail.

Finally, it should be noted that differences in velocities
(Table~\ref{t:lorre_evf}) are much smaller than what is usually
expected in the cases of dynamical families produced by
disruption events. In fact, Lorre seems to be likely
issued from a moderate-energy cratering event, and is the most
compact group known so far among high-inclination families
\citep{Novakovic2011}.

\subsubsection{Size-frequency distribution}

Some important information about the impact physics can be obtained
by studying the size-frequency distributions (SFDs) of asteroid
families \citep{Tangaetal99, durda2007}. It is generally found that
these distribution can be described by a power law, N($>$D) $ \propto D^{- \alpha}$. 
Younger asteroid families generally have steeper SFDs 
which are generally thought to evolve with time toward shallower 
trends due to collisional and dynamical erosion of the family. 
A correct way to fit these distributions, i.e. to estimate exponent $\alpha$, is to adopt an 
approach based on maximum likelihood method applied to bi-truncated 
Pareto distributions \citep[see][]{cellinoetal91,Tangaetal99}.
However, the number of family members is currently too small to 
perform such a statistical analysis.

Thus, we used an alternative approach based on the least-squares method\footnote{This 
approach, despite being widely used, is not correct strictly speaking. In particular,
this method may significantly underestimate the uncertainties of obtained values. However, as 
in any case the number of known members is too small to obtain a highly reliable result, we 
used this method because of simplicity of its implementation.} to estimate the exponent $\alpha$.
In this way, by fitting cumulative size distribution, for objects between 3.0 and 
4.5 km in diameters, we found $\alpha$ to be $3.2$. This value
is smaller than expected for typical young asteroid families \citep{karin2006,parker2008}.
Likely, this result is affected by the observational incompleteness
and a real $\alpha$ is somewhat larger.
In any case, for a moment, we can only say qualitatively that the
cumulative size distribution does not appear to be very steep.

\begin{figure}
\includegraphics[angle=-90,scale=.30]{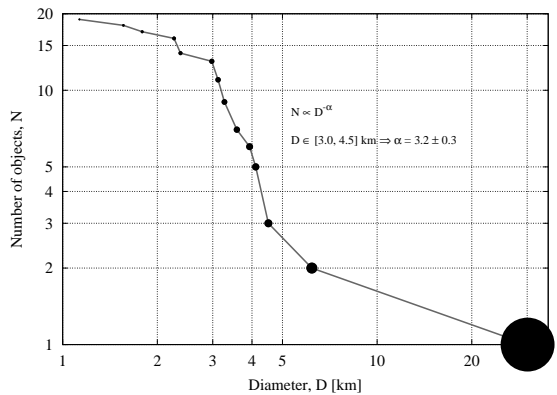}
\caption{Cumulative size distribution N($>$D) of 19 asteroids belonging to the Lorre cluster.}
\label{f:sfd}
\end{figure}

\subsubsection{Collisional lifetime}

It is interesting to estimate what was expected collisional lifetime of
the Lorre cluster parent body. This computation depends on many parameters,
including mainly the inventory and size distribution of the possible
impactors, the average impact velocity and, for what concerns the
outcomes of the collisions, on the impact strength of the body,
which in turn depends on its size and density.

We computed the mean intrinsic collision probability and the
mean impact velocity for the collisions between (5438) Lorre
and other main-belt asteroids using the approach of \citet{aldo1998}. 
The mean impact velocity results to be about $10$ km~s$^{-1}$, due to
the high-inclination orbit of (5438) Lorre. Under standard assumptions
on the cumulative size distribution of the population of possible projectiles,
described by a power-law with an exponent of $2.5$, a 
density value of $1.5$ g~cm$^{-3}$, and setting the impact strength on the basis of the 
results of \citet{benz1999}, the estimated collisional lifetime of (5438) Lorre is
$6.6$~Gyr, in agreement with results of some independent studies \citep{bottke2005}.
This relatively high value does not change much by steepening the size
distribution of the projectiles (the lifetime becomes $5.3$~Gyr if
the power-law exponent is increased to the value of $3.0$), nor by changing the
value of the density.

Asteroid Lorre is isolated, and there are no asteroids of
similar size in its surroundings which might have been produced by
the disruption of a hypothetical common parent body. We are led
therefore to conclude that Lorre could be a pristine
asteroid, which survived nearly intact since the time of its
formation. This makes its analysis even more interesting.

\section{Summary and Conclusions}
\label{s:conclusions}

Here we show the first example of a young asteroid cluster located in a dynamically stable
region, which was produced by partial disruption of a primitive body about 30 km in size. We
estimate its age to be only $1.9\pm0.3$ Myr, thus its
post-impact evolution is very limited. The large
difference in size between the largest object and the other cluster
members means that this was a cratering event. The parent body had a
large orbital inclination, and was subject to collisions with
typical impact speeds higher by a factor of $2$ than in the most
common situations encountered in the main belt. For the first time
we have at disposal the observable outcomes of a very recent event
to study high-speed collisions involving primitive asteroids,
providing very useful constraints to numerical simulations of these
events \citep{michel2003,jutzi2009,leinhardt2012} and to laboratory
experiments \citep{housen2011}.

This is the best preserved young asteroid family produced by partial disruption of
a primitive asteroid, of a kind which is supposed to have survived nearly unaltered
since the epoch of formation of the Solar System. Being young and well distinct
from the background population, this cluster provides very useful information that
can help to answer several long-debated questions in planetary science. Examples
include a better understanding of impact physics, material strength and the role of
space weathering. These process, highly dependent on the composition of
the objects, are so far poorly constrained for primitive asteroids.

Among the members of the Lorre cluster there are several asteroid
pairs, couples of objects with nearly identical orbital parameters.
These pairs may well consist of couples of fragments which were
ejected with nearly identical ejection velocities. Another
possibility is that they might actually be the components of former
binary systems originally produced by the collision, and later
decoupled by some mechanisms \citep{pravec2010}.
Production of binary systems in collisional events has been
suggested by numerical simulations \citep{michel2001,durda2004}, but their
expected abundance in asteroid families has not been firmly
established yet. The young age of the Lorre cluster
as well as its sharp separation from background objects may potentially
help to better understand both populations, binaries and pairs.

An interesting possibility for future work comes from a recent result of 
\citet{benavidez2012} how found that low-energy impacts into rubble-pile 
and monolithic targets produce different features in the
resulting SFD, and, thus, this is a potentially diagnostic tool to
study the initial conditions just after the impact and the internal
structure of the parent bodies of asteroid families. According to \citet{benavidez2012},
cratering events, produced by small impactors, can potentially provide even 
more information about the internal structure of the parent body than
catastrophic or super-catastrophic events produced by large impactors.
Thus, the Lorre cluster seems to be a very promising candidate.

Next, the Lorre cluster may be very useful to improve
our knowledge about space weathering processes acting on primitive
bodies, a debated subject since results based on the Sloan
Digital Sky Survey broadband photometry \citep{Nesvorny2005} are not
consistent with the results of some laboratory experiments \citep{brunetto2009}.

Finally, the cluster may be a very interesting place to search for new main-belt comets (MBCs).
\footnote{Main belt comets are objects dynamically indistinguishable from main belt asteroids,
but which exhibit comet-like activity due to the sublimation of volatile ice \citep{Hsieh2006}.}
A recent findings by \citet{nov2012} supports an idea that this kind of objects
may be preferentially found among the members of young asteroid families \citep{Nesvorny2008,Hsieh2009}.
In this respect, members of Lorre cluster are particularly interesting candidates because their
heliocentric distances are smaller than those of currently known MBCs. Thus,
they may provide a clue about the inner edge of populations of MBCs.

\section*{Acknowledgements}
The work of B.N. and Z.K. has been supported by the Ministry of
Education and Science of Republic of Serbia under the Project 176011.
B.N. also acknowledges support by the European Science Foundation
through GREAT Exchange Grant No. 3535.

\end{document}